# COMPUTATION OF INTERNAL FLUID FLOWS IN CHANNELS USING THE PACKAGE CFX-5

**ALEXEY N. KOCHEVSKY**
*Research Scientist, Department of Applied Fluid Mechanics,*
*Sumy State University,*
*Rimsky-Korsakov str., 2, 40007, Sumy, Ukraine*
*alkochevsky@mail.ru*

**Abstract:** The paper presents the results of computation of some simple internal flows of incompressible fluid with the software package CFX-5. In the previous paper of the author, the same flows were used for testing of another CFD software tool, FlowVision. The model equations used for this research are again the set of Reynolds and continuity equations and equations of the standard $k - e$ turbulence model. The results of computation of a number of internal fluid flows in channels are presented and comparison with known experimental data is performed. Good correspondence of results was obtained both concerning the flow pattern and related integral values.

## Introduction

Tendency of recent years is appearance and wide distribution of commercial CFD software tools allowing for performing of numerical simulation of fluid flows of arbitrary complexity in the regions of arbitrary geometrical configuration. However, in order to prove reliability of these tools for different flows, profound comparison with known experimental results is required.

Description of approach for simulation of fluid flows using modern CFD software tools is presented, e.g., in the papers of Kochevsky (2004). The present paper deals with the software package CFX-5. This package was granted for test application to the Sumy State University, and in this paper we present the results of our testing, by comparing the computational results obtained with this tool with known experimental data for a number of simple fluid flows in channels. In the previous paper of the author (Kochevsky, 2004a), the same flows were used for testing of another CFD software tool, FlowVision (www.flowvision.ru). The present paper considers the problems for which the discrepancies between the numerical and experimental results in the previous paper were the most significant. Thus, the present paper permits to compare the capabilities and precision of these two packages at an example of the considered flows.

## CFX: general information and history of development

The abbreviation “CFX” is decoded as “Computational Fluid dynamiX”.

Predecessors of the package CFX-5 (the last version of which, CFX-5.7, license No. 13200000, was used for this research) were the packages TASCflow and CFX-4. The former was developed in Waterloo, Canada, and for many years was among the world leaders in numerical simulation of flows in turbomachinery. The latter was developed in England, its advantage was capability for modeling of complex flows in chemical industry. Several years ago, AEA Technology, an English company, has become the master of both these packages and created on their basis a new package, CFX-5 (mostly on the basis of TASCflow) that has combined the advantages of both predecessors. The package was also improved by the German group of developers that dealt with development and testing of advanced turbulent models. At last, the company ANSYS has become recently the owner of the package. The well-known package ANSYS produced by this company is during many years one of the world leaders in all types of engineering simulations, especially in strength analysis. At present, the package CFX-5 is a part of the package ANSYS for simulation of fluid flows, and is commercially available also separately.

A lot of information regarding the package CFX-5 and its capabilities can be found at the site www-waterloo.ansys.com/cfx/ and at the regional sites of developers and distributors. In particular, new features of the last versions of the package and examples of successful simulation of complex flows are described in the periodical journal CFX Update, the full texts of which are available for downloading at this site.

Among a number of flow models implemented in the package CFX-5, in this paper, like in the paper of Kochevsky (2004a) devoted for FlowVision, we used only the model of turbulent flow of incompressible fluid. $K - e$ turbulence model with scalable wall functions was used (Kochevsky, 2004b).

## Results

### 1. Flow in diffusers of large internal angles

An advantage of the package CFX-5, like also FlowVision, is capability to compute successfully flows with symmetrical computational domain and symmetrical boundary conditions, where the flow is not symmetrical and, moreover, transient.

So, in a conical diffuser with large internal angle the flow separates and attaches to part of the wall in random way. The position of separation zone varies randomly along circumference of diffuser (Idelchik, 1975). This known physical phenomenon is reproduced by computation (Fig. 1). Though, according to the results of computation in CFX-5, for this internal angle, the flow separates along the whole circumference of the walls and looks like a jet.

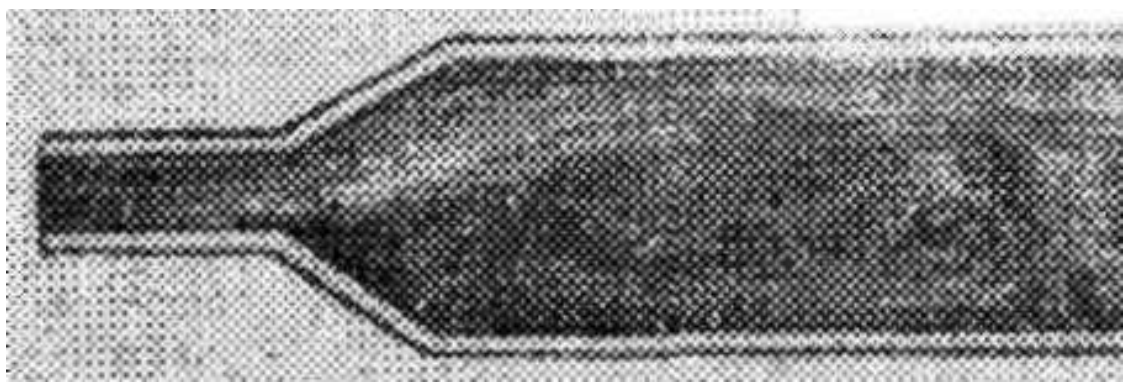

a) experiment,
Idelchik (1975)

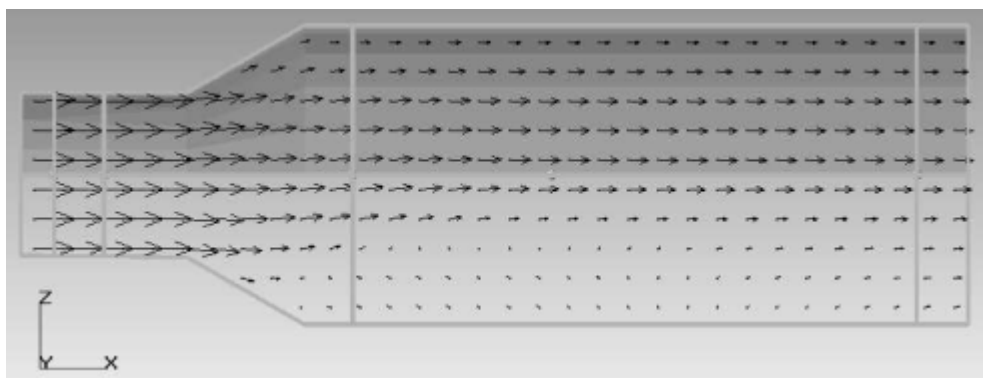

b) computation, FlowVision,
Kochevsky (2004a)

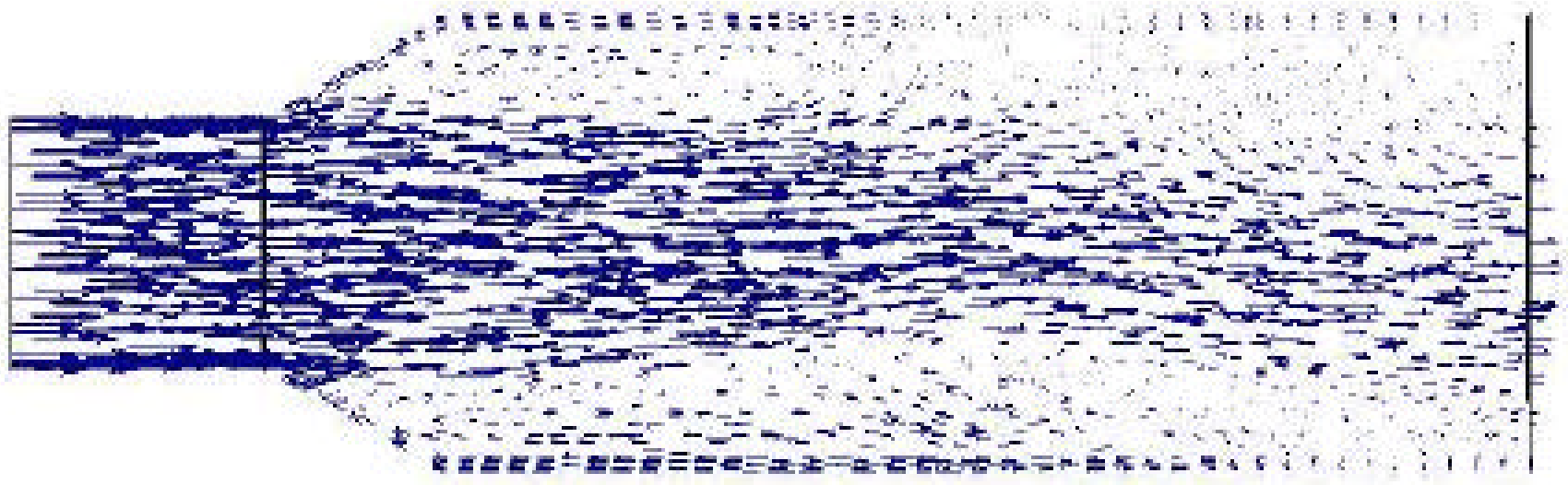

c) computation, CFX-5

**Figure 1.** Instant flow pattern in a conical diffuser with area ratio 3.3 and internal angle 60°

Figure 2 presents corresponding loss factors obtained by computation in FlowVision and CFX-5 and taken using interpolation from that reference book of Idelchik (1975). Note that for this flow the solution process does not converge to the stationary solution, and the computed loss factors presented at Fig. 2 are approximate time-averaged values.

For numerical simulation in CFX-5, a computational mesh of about 100 000 cells was used. As it is seen, the results of computation in CFX-5 are much closer to the experimental values in comparison with FlowVision. It should be also noted that at the outlet of the computational domain shown at the Fig. 1c, the velocity distribution is still far from the logarithmic shape. Therefore, further smoothing of the flow downstream will cause additional losses. Taking this into account, the correspondence of numerical and experimental results may be considered as good.

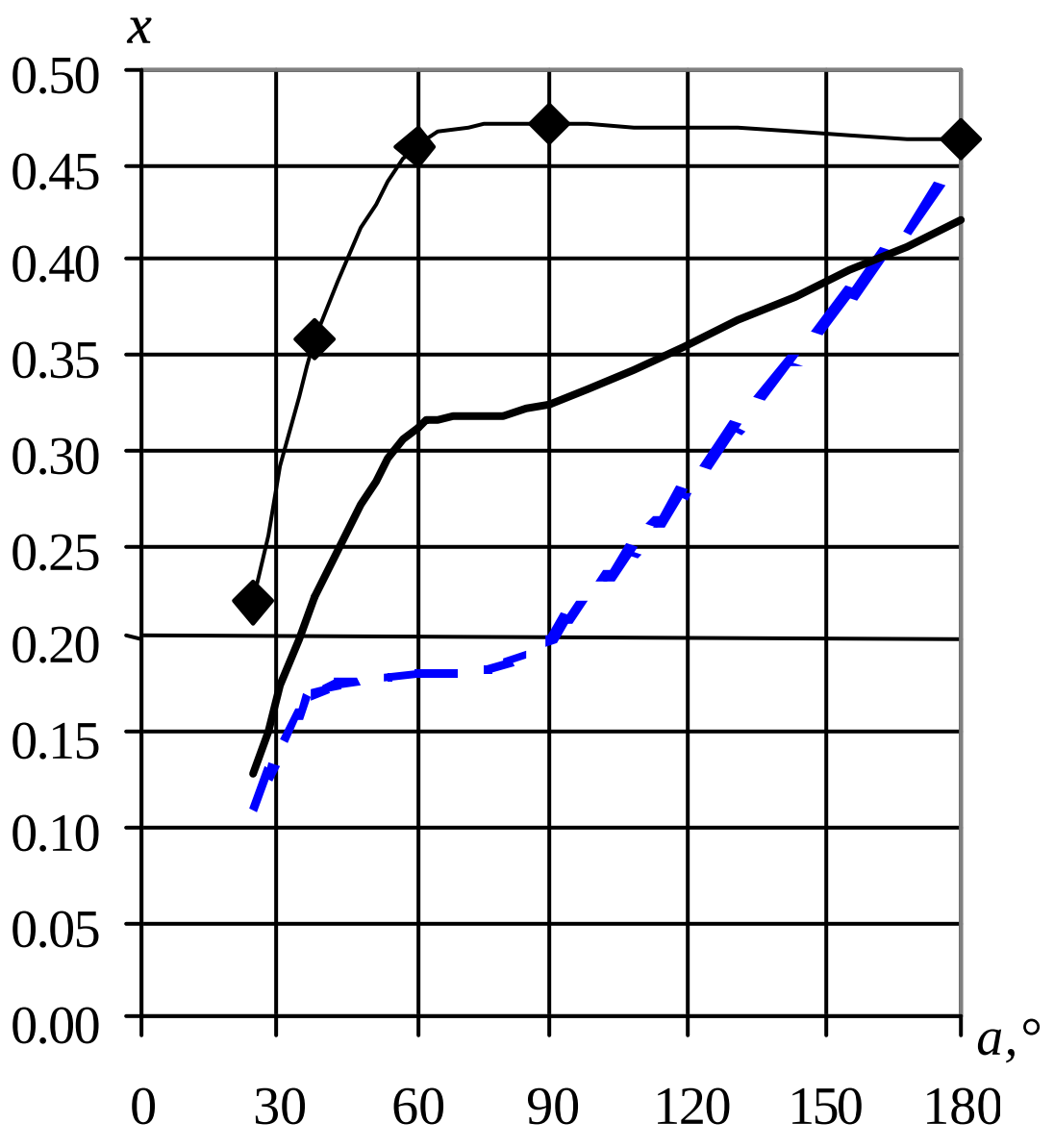

**Figure 2.** Dependence of loss factor on internal angle of a conical diffuser with area ratio 3.3: markers connected with the thin line – experiment, Idelchik (1975), dotted line – computation, FlowVision (Kochevsky, 2004a), solid line – computation, CFX-5

**2. Flow in a 180-degree bend of square cross-section**

For comparison, experimental results of Chang et al. (1983) are used. The considered channel is presented at Fig. 3. The flow is symmetrical relative to the medium plane of the channel, thus, in order to spare computational resources, only half of the channel was used as the computational domain. Reynolds number computed by hydraulic diameter was 58000.

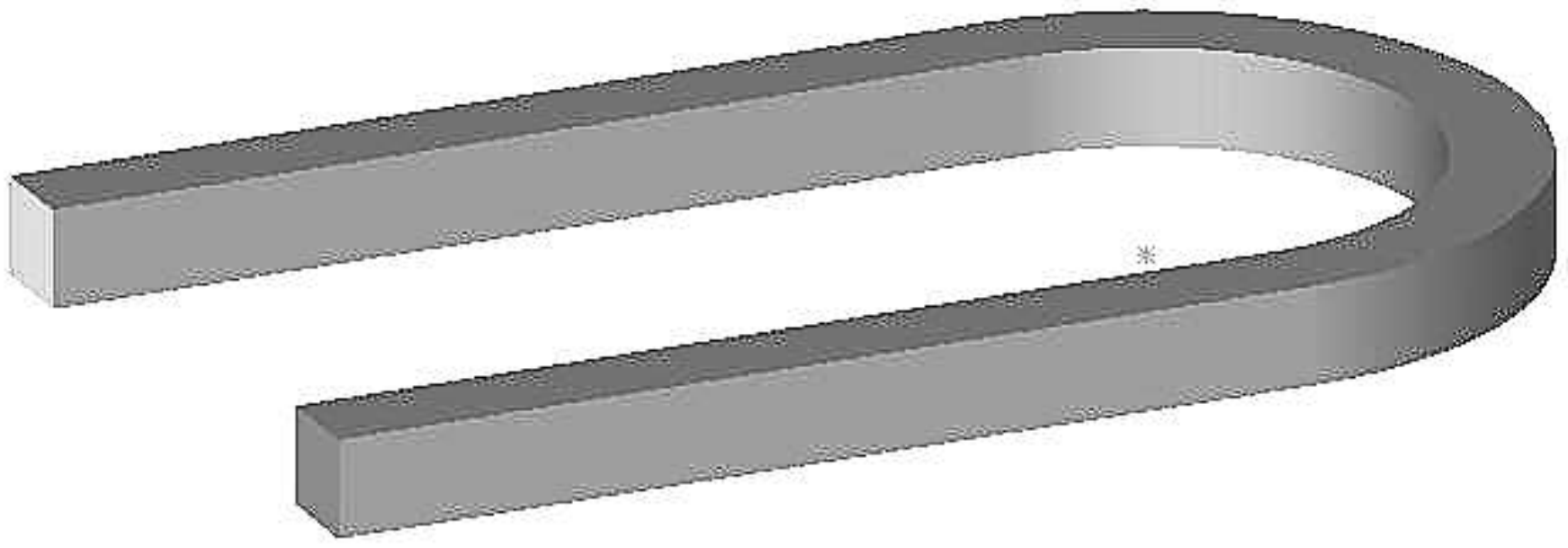

**Figure 3.** Geometrical configuration of the channel

A physical effect peculiar to this flow is the twin vortex occurring as the flow turns (Fig. 4). Fluid moves to the external radius of bend along the symmetry plane of the channel and returns back along the side walls. Figure 5 presents distributions of axial velocity at several intermediate cross-sections of the bend ($H$ is the height of a cross-section of the channel). The distributions are taken at the plane equidistant from the symmetry plane and the lower wall of the channel.

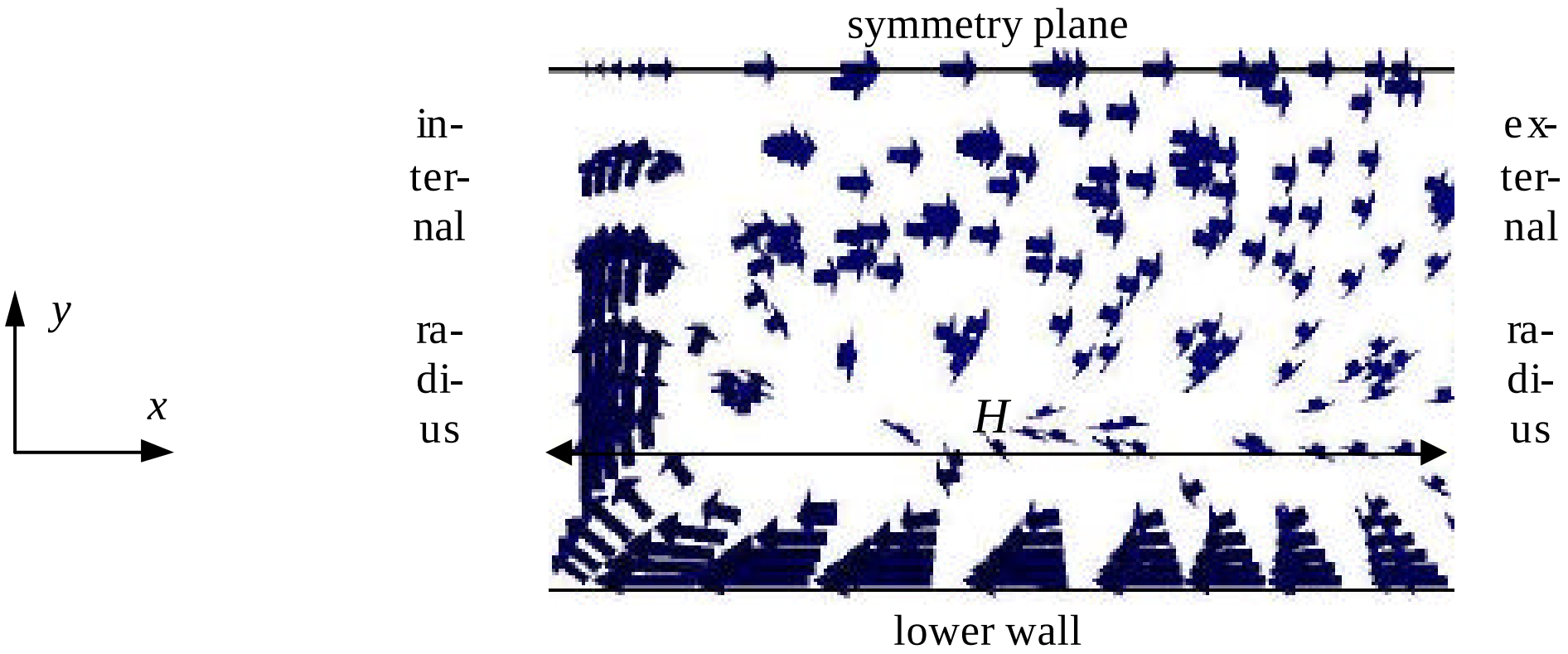

**Figure 4.** Velocity vectors at the cross-section corresponding to 90-degree angle of turn

For conduction of this research in CFX-5, a computational mesh with about 170 000 cells was used. It is seen that the results of computation using CFX-5 coincide with the experimental data much better that the results of computation with FlowVision. A probable reason for superior prediction is the layout of the computational mesh in CFX-5 that suits to the shape of the channel and provides for good resolution of the boundary layer (in contrast to the Cartesian computation mesh used in FlowVision). Another reason is more efficient solver of CFX-5 that permits to use a mesh with much larger number of cells for the same computational resources of a PC. Finally, implementation of scaled wall functions for the $k - e$ turbulence model is an additional reason for improved precision of predictions in CFX-5.

As one can see, the results computed with CFX-5 coincide with the experimental data quite well, however, some quantitative divergence is observed. In particular, low velocities at the middle of the graph in the cross-sections of 90° and 130° were observed at the experiment but were not predicted by the computation.

This flow was investigated numerically also in the paper of Choi et al. (1989) using a similar computational algorithm but more complex turbulence models. The conclusion was made that for this flow, the standard $k - e$ turbulence model gives somewhat inaccurate solution, more precise results were obtained, in particular, using the Reynolds algebraic stress model.

### 3. Flow in a rotating channel

For comparison, the experimental results of Moore (1973) are used. Geometrical configuration of the channel is presented at Fig. 6. As one can see, it resembles in many aspects the geometrical configuration of a blade-to-blade channel of a radial-flow impeller with radial blades. The channel rotates around the axis of the inlet cylindrical section. The flow enters the channel along its rotation axis, then turns in radial direction and expands. Outlet section of the channel is open to the atmosphere.

The flow enters the channel through a stabilizing device ensuring the through-flow velocity is constant through the inlet cross-section. Then channel walls gradually involve the flow in rotation. Rotational speed of the channel was 206 rpm. Upstream of the stabilizing device, a blower was provided, allowing for supply of air to the channel with different flow rates.

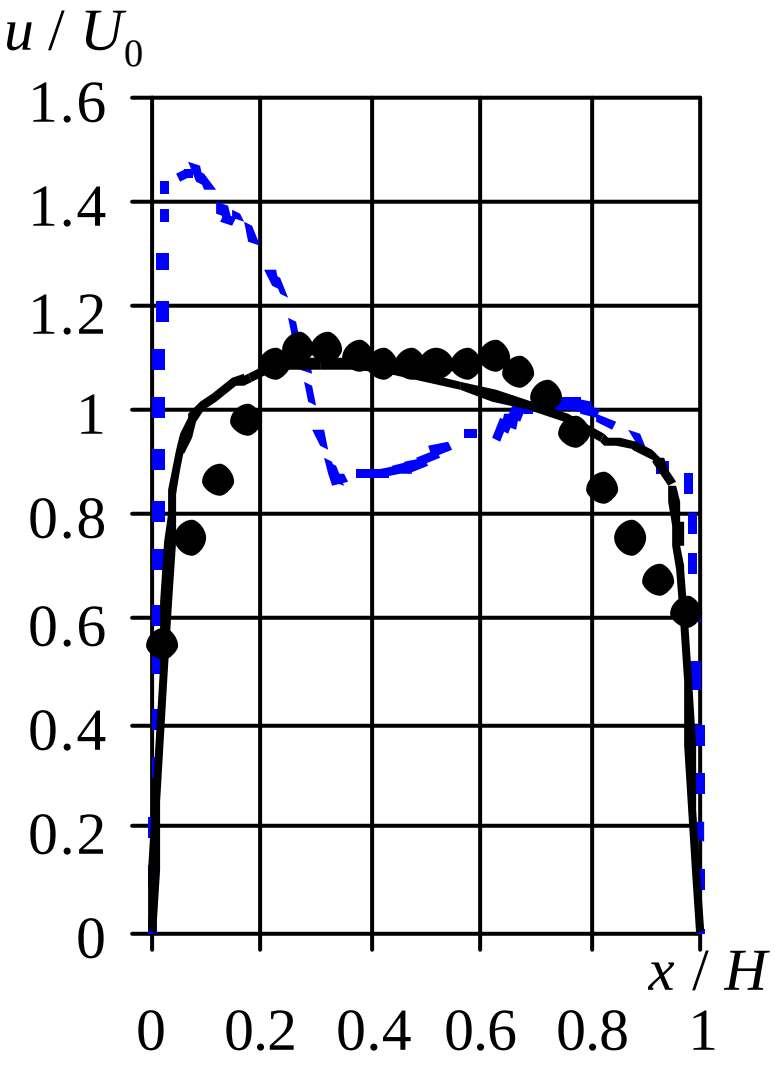

a)

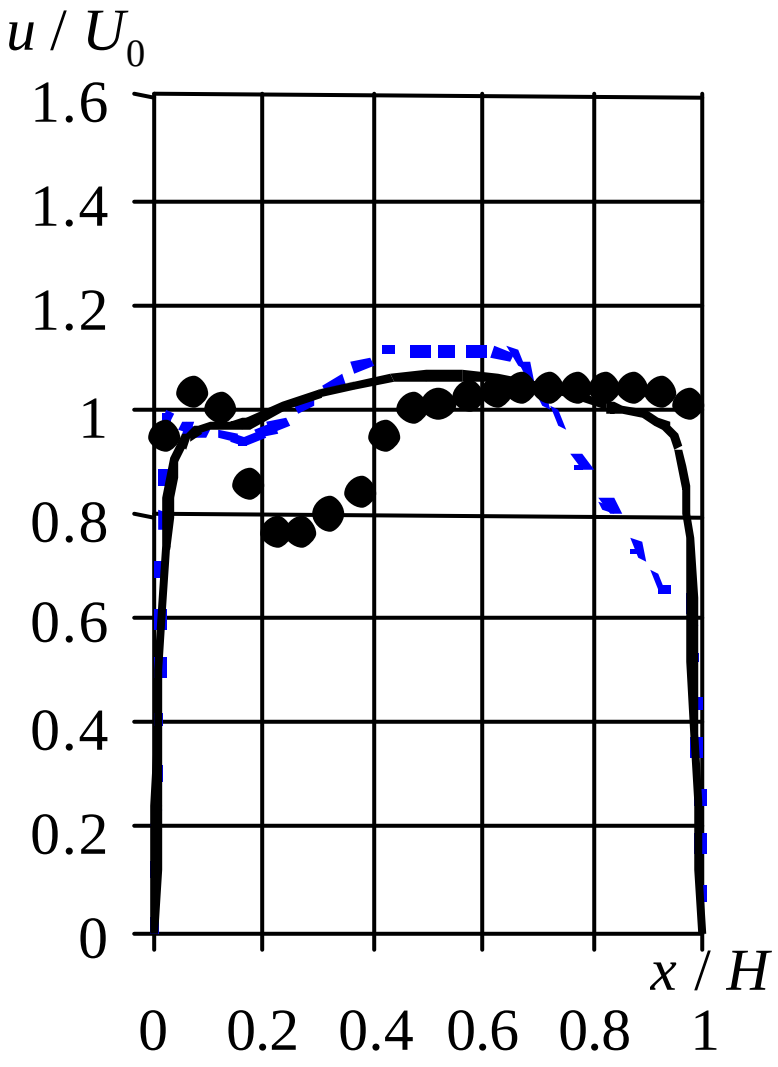

b)

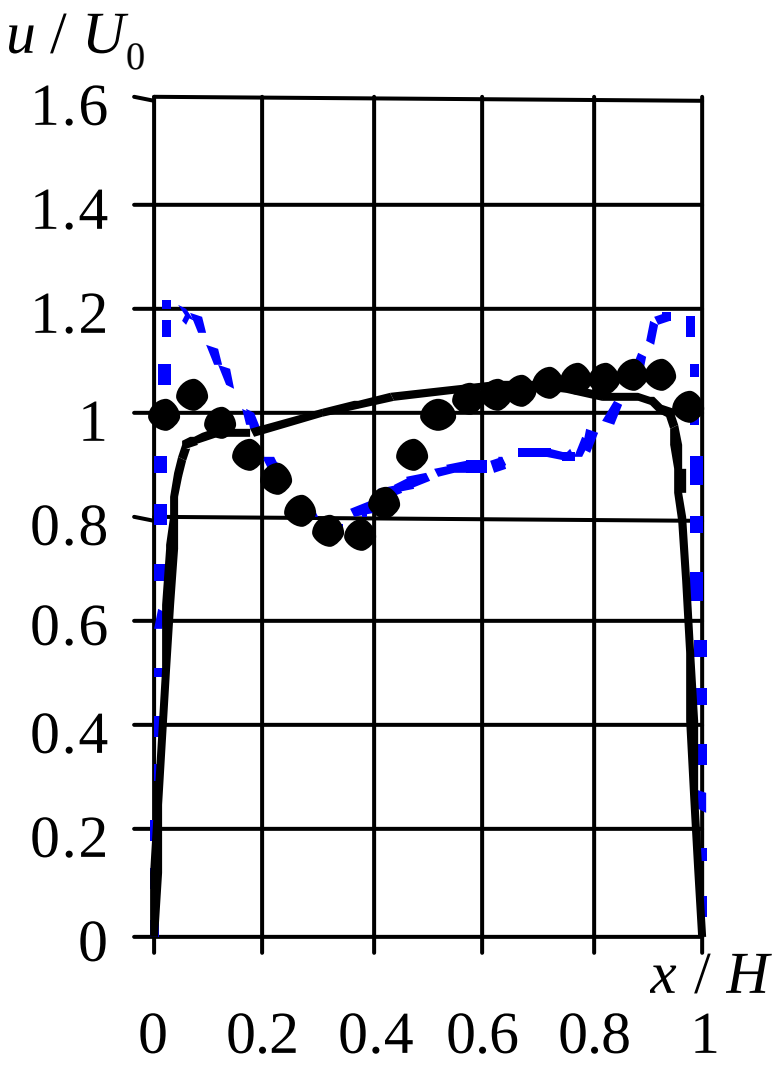

c)

a) Angle of turn 45°
b) Angle of turn 90°
c) Angle of turn 130°

**Figure 5.** Axial velocity distributions along the bend ($y / H = 0.25$): markers – experiment, Idelchik (1975), dotted lines – computation, FlowVision (Kochevsky, 2004), solid lines – computation, CFX-5

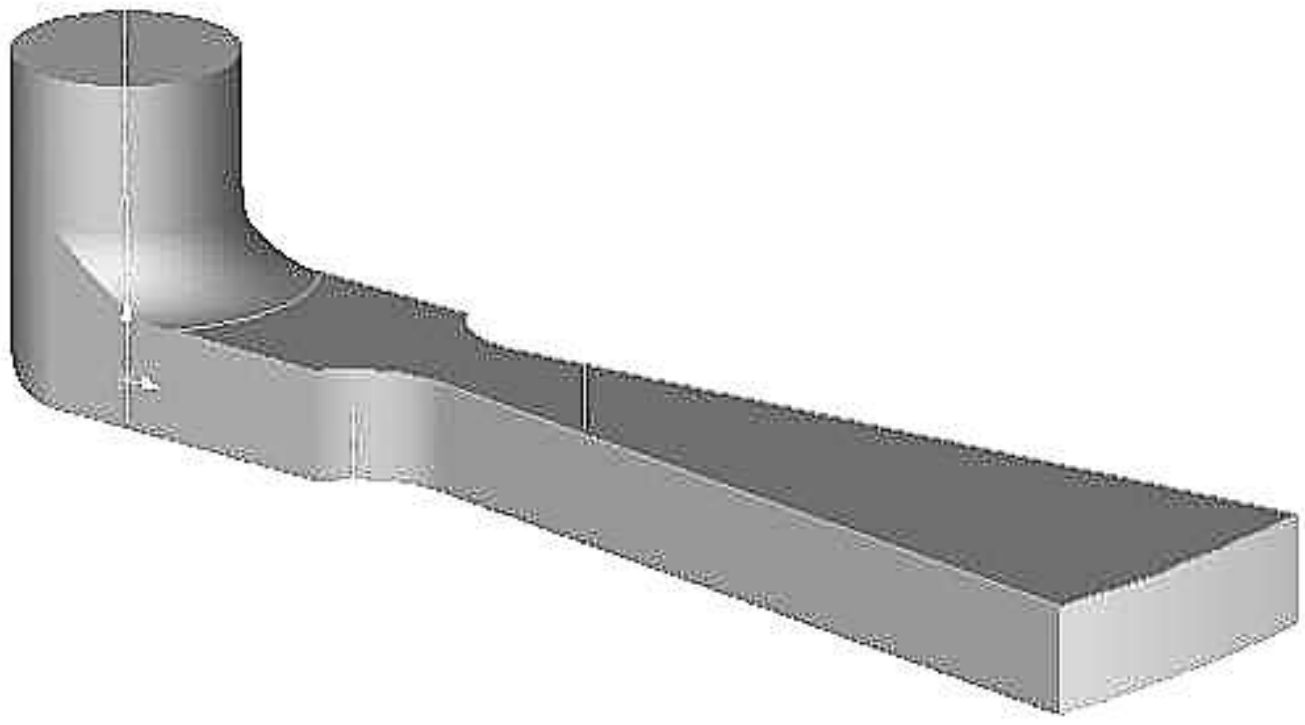

**Figure 6.** Geometrical configuration of the channel

This flow is featured with the following physical effects.

- As distance from the rotation axis increases, static pressure grows rapidly due to action of centrifugal force. Additionally, it grows due to expansion of the channel.
- The pressure $p$ is larger at the pressure side than at the suction side. As it follows from the analysis of potential flow (not regarding viscous effects), the pressure decreases from the pressure side to the suction side, according to the linear law $\partial p / r\, \partial y = -2\, w\, u$, where $w$ is angular rotational speed of the channel.
- As it follows from the analysis of potential flow, through-flow velocity $u$ increases in every cross-section from the pressure side to the suction side, according to the law $\partial u / \partial y = 2\, w$ (Fig. 7).
- At low flow rate, the flow in the channel is pressed to the suction side along the whole length of the channel. Thus, at the pressure side, the reverse flow is observed when the distance from the rotation axis is large enough (Fig. 7).

All these effects are well reproduced by computation.

- According to the experiment, when the flow rate is large enough, the flow at some distance from the rotation axis separates from the suction side and is pressed to the pressure side. The reason for this is seemingly rapid growth of pressure along the suction side as the channel expands (Fig. 8). Just near this wall the flow is inclined to separation. At low flow rate this separation does not occur, because earlier the flow separates from the pressure side.

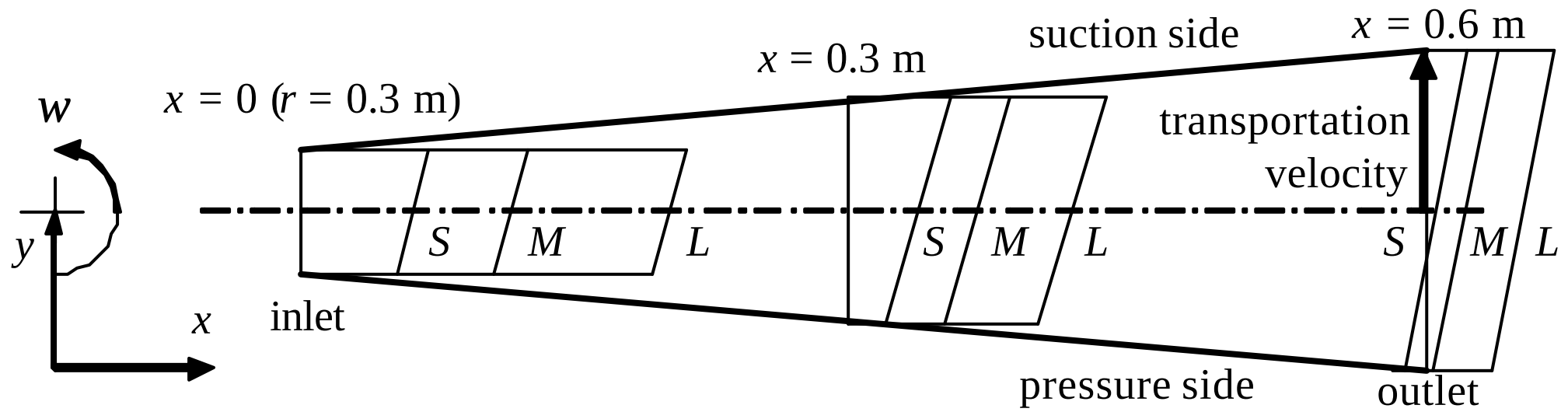

**Figure 7.** Velocity distribution in the channel according to the potential flow analysis: S – low flow rate (average velocity at the outlet cross-section of the channel is 2.8 m/s), M – medium flow rate (5.3 m/s), L – high flow rate (11.0 m/s)

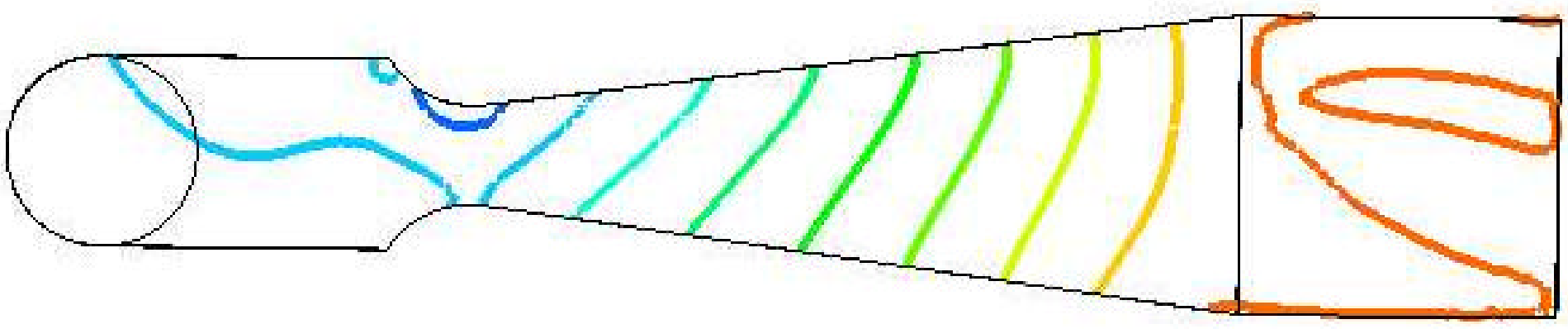

**Figure 8.** Isolines of static pressure at the middle-height cross-section of the channel at medium flow rate, computation, CFX-5

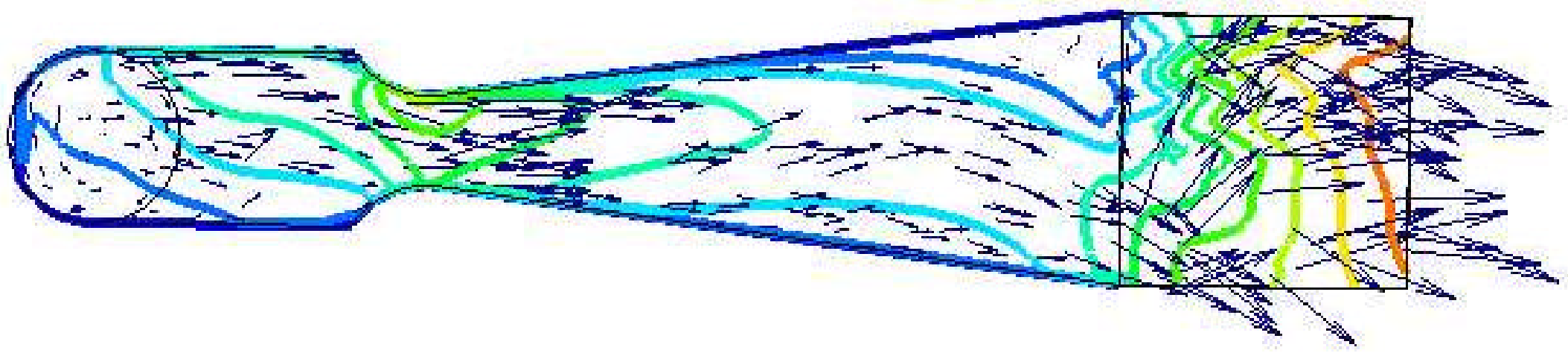

**Figure 9.** Velocity isolines and vectors in relative motion at medium flow rate, computation, CFX-5. At the upper right corner, near the channel outlet, a stagnation zone is seen

This rather simple experiment is a good illustration explaining why location of separation zones in blade-to-blade channels of a radial-flow impeller depends significantly on the flow rate.

For performing computation in CFX-5, we have used the computational grid containing 120 000 cells. The computation was performed in the rotating frame of reference. Here, as the inlet boundary condition, in addition to through-flow velocity, solid body swirl was imposed. The rotational speed of the swirl was – 206 rpm (opposite to the direction of channel rotation). In order to simulate properly the flow at the outlet region, the computational domain included also the space downstream of the channel, where the flow went to the atmosphere (Fig. 8 – 10). Figure 10 shows vectors of flow velocity: at the inlet region – in relative motion, at the outlet region – in absolute motion.

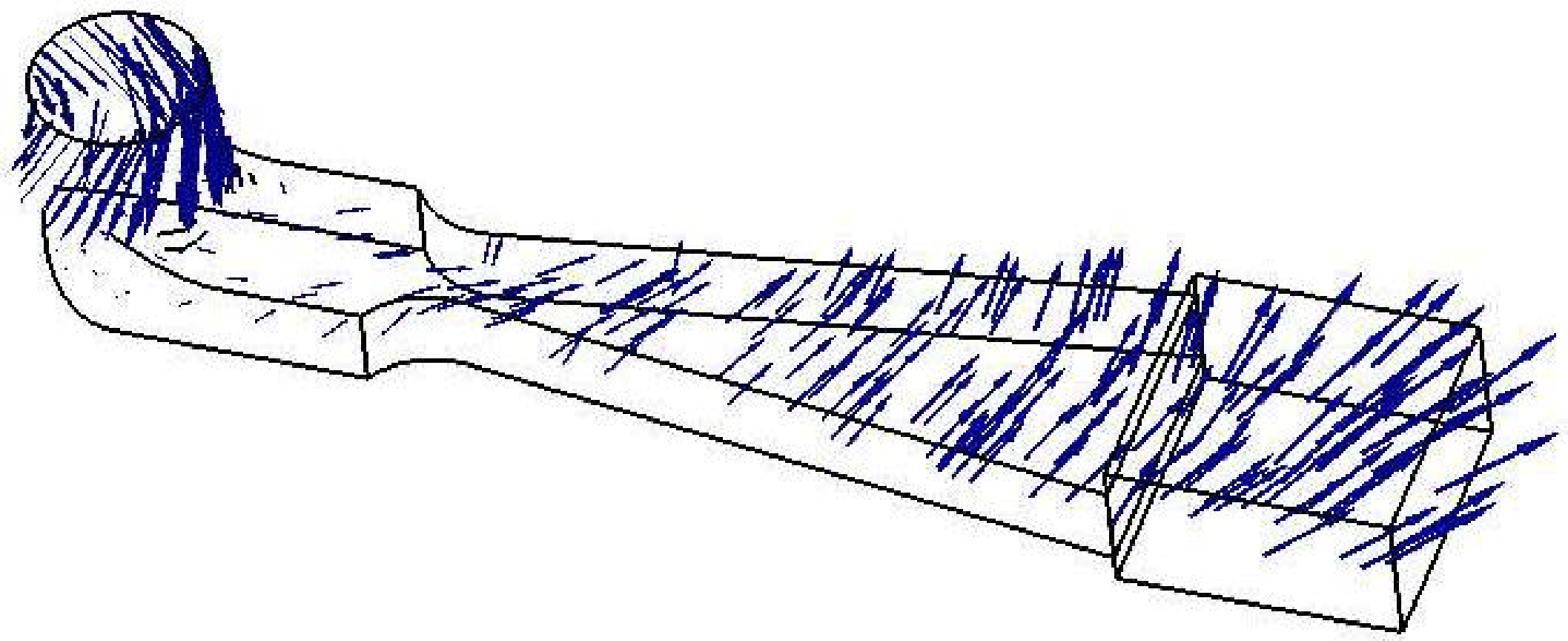

**Figure 10.** Velocity vectors of flow in relative motion (at the channel inlet) and in absolute motion (inside the channel and at the channel outlet) at medium flow rate

Note that in the paper of Majumdar et al. (1977) this flow was computed using a sweep scheme, with the same $k-e$ turbulence model. The separation of flow from the suction side was not observed in that paper, thus demonstrating that Reynolds equations should not be simplified by order-of-magnitude analysis for simulation of this flow.

Figure 11 presents the computed and experimental velocity distributions at the cross-section before channel outlet, at different flow rates. It is seen that the results of computation in CFX-5 correspond to the experimental results better than the results of FlowVision. Nevertheless, quantitative discrepancy between the results is still available. The reasons for this discrepancy are seemingly the same as for the previous flow.

## Conclusion

For the flows presented here, we have obtained good qualitative and at least satisfactory quantitative correspondence of the results obtained using the software package CFX-5 with the experimental results. For types of flows presented here, we should recognize the CFX-5 to be a good tool for numerical simulations and more precise in comparison with FlowVision.

### *Acknowledgements*

The present research was conducted under leadership of Asst. Prof. A.A. Yevtushenko and under support of the collective of the department of applied fluid mechanics.

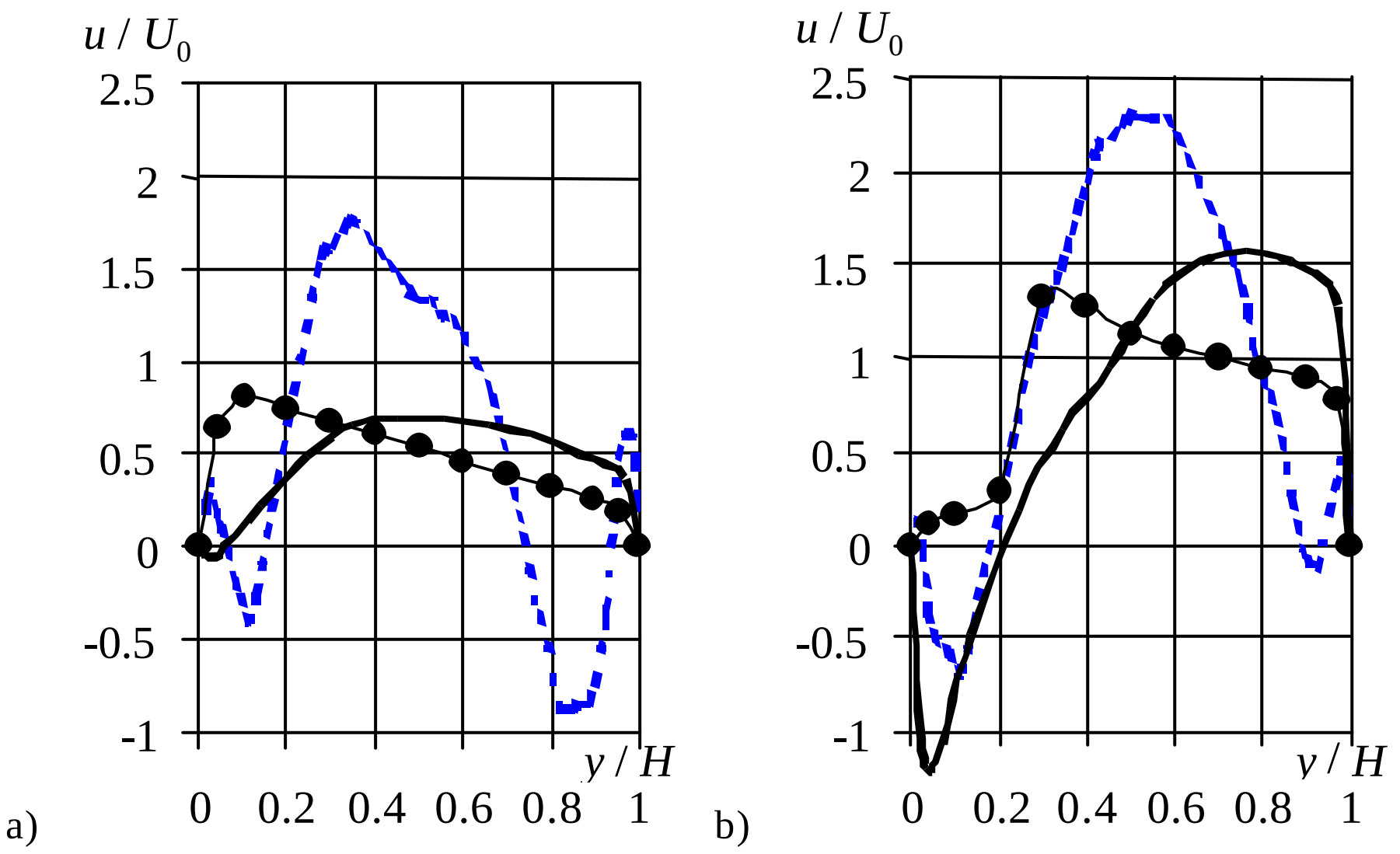

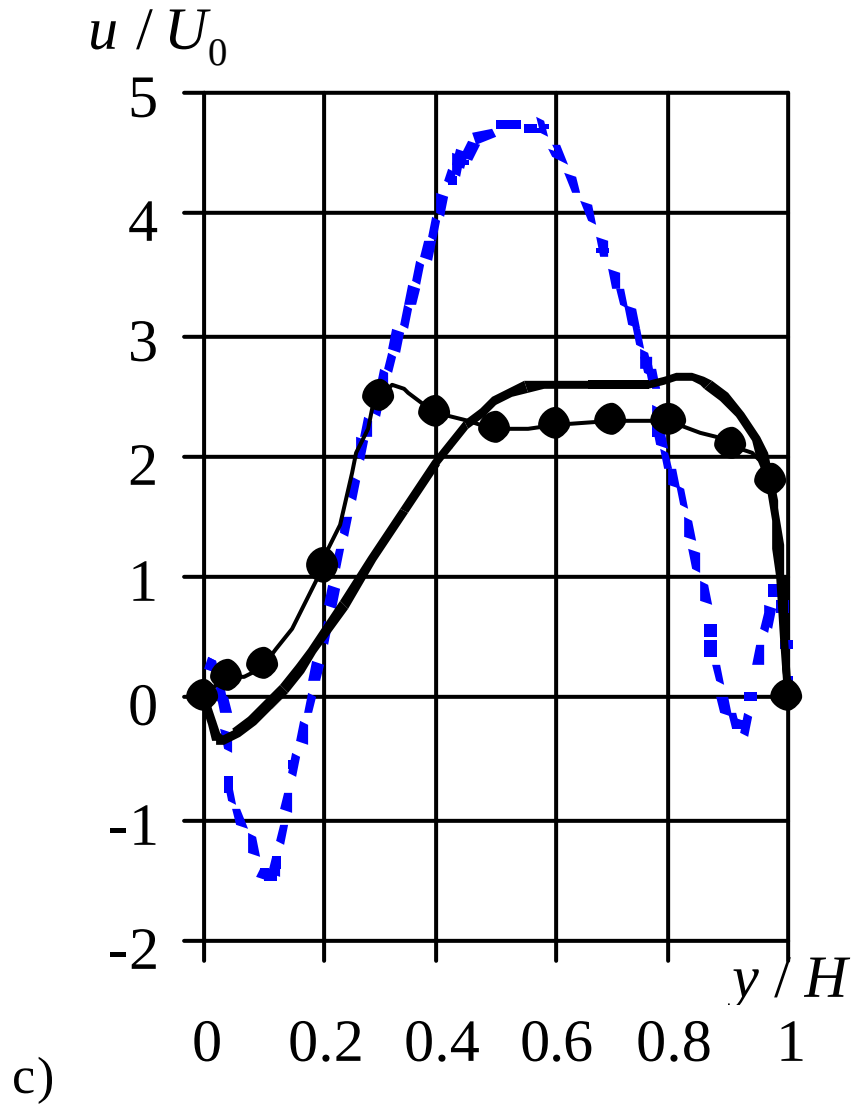

a) Low flow rate
b) Medium flow rate
c) High flow rate

**Figure 11.** Distributions of through-flow velocity at the cross-section $x$ = 0.46 m: markers connected with the thin lines – experiment, Moore (1973), dotted lines – computation, FlowVision (Kochevsky, 2004), solid lines – computation, CFX-5. Left-hand side is the suction side. $U_0$ – average through-flow velocity at the channel inlet, at the medium flow rate, $H$ – height of the channel cross-section